\documentclass[aps,prd,reprint,superscriptaddress,amsmath,amssymb]{revtex4-2}

\usepackage{graphicx}% Include figure files
\usepackage{dcolumn}% Align table columns on decimal point
\usepackage{bm}% bold math
\usepackage{xcolor}
\usepackage{amssymb}
\usepackage{lipsum}
\usepackage{amssymb,amsmath}
\usepackage{amsthm}
 \usepackage{mathtools}
\usepackage{physics}
\usepackage{graphicx}% Include figure files
\usepackage{tikz}
\usepackage{physics}
\usepackage{tikz}
\usepackage{MnSymbol}
\usepackage{feynmp}
\usepackage{float}
\usepackage{hyperref}
\usepackage{longtable}
\usepackage[T1]{fontenc}

\begin{document}

\preprint{APS/123-QED}

\title{Quantum Scattering in Schwarzschild Spacetime:\\ Hawking Radiation and Black Hole Atmospheres}% Force line breaks with \\%u

\author{Victor H. Alencar}
 \email{victoralencar@pos.if.ufrj.br}
 \author{Gabriel Picanço}
 \email{g.picanco@if.ufrj.br}
\author{Carlos A. D. Zarro}
 \email{carlos.zarro@if.ufrj.br}
\affiliation{%
Instituto de Física - Universidade Federal do Rio de Janeiro\\
 Rua Milton Santos, 117 Cidade Universitária, CEP: 21941-585, Rio de Janeiro - RJ, Brasil}%



\begin{abstract}
In this paper, we investigate scattering of a scalar field near a Schwarzschild black hole through its $S$-matrix. Within this framework, we obtain a novel derivation of Hawking radiation by computing the emission rate, which yields a Bose--Einstein distribution with temperature $T_H=(8\pi GM)^{-1}$, the Hawking temperature. In addition to Hawking radiation, the S-matrix exhibits antibound states, corresponding to excitations at the threshold of becoming scattering (bound) states if the potential is decreased (increased). We interpret these excitations as constituents of the black-hole quantum atmosphere: a thermalised region outside of the event horizon, which is the source of the Hawking radiation. Using the spectrum of antibound states, we found the atmospheric radius, $r_{\text{Atm}} \approx 2.77 r_{s}$, which is in good agreement with previous results in the literature obtained through other methods. Our results indicate that, at the macroscopic level, the quantum atmosphere behaves like an ordinary thermalised gas at the Hawking temperature.
\end{abstract}

\keywords{Black Hole Thermodynamics, Quantum Scattering in Curved Spacetimes, Black Hole Atmospheres, S-Matrix}
\maketitle

%\tableofcontents

\section{Introduction}

Currently there is no consistent method to quantise gravity, as the perturbative renormalisation program \cite{Schwartz:2014sze} that successfully describes the electromagnetic and nuclear forces fails when applied to general relativity \cite{tHooft:1974toh, Deser:1974cy, Deser:1974cz, Deser:1974xq, Goroff:1985th, Percacci:2017fkn}. Even so, remarkable breakthroughs have been made by considering quantum fields in classical curved spacetimes \cite{Birrell:1982ix, Wald:1995yp, Witten:2024upt}, such as the entropy of black holes and the Hawking radiation \cite{Hawking:1975vcx, Parikh:1999mf}, \textit{i.e.} the prediction that, at the quantum level, black holes emit thermal radiation due to quantum fluctuations.

When the Hawking radiation was first derived \cite{Hawking:1975vcx}, it was believed that the emission was from the event horizon, which in the Schwarzschild case is a null $2-$sphere of radius $r_{s}=2GM/c^2$. Subsequent investigations predicted the existence of black-hole quantum atmospheres \cite{Giddings:2015uzr, Hod:2016hdd, Dey:2017yez, Dey:2019ugf, Ong:2020hti, Gingrich:2023qae, Kaczmarek:2024QuantumAtmosphere, Zhang:2026ixo, Liu:2026exs}. In this picture, the Hawking radiation is not emitted by the event horizon but from a surface with radius $r_{\text{Atm}}>r_{s}$. The volume between the two radii is the black hole atmosphere, a thermalised region at $T_{H}=\frac{1}{8\pi G M}$, the Hawking temperature.

Predating the results on quantum aspects of black holes, scattering theory on such spacetimes played a central role in black-hole physics. Mainly focusing on classical aspects of scattering, such as the response of black holes to external perturbations \cite{Regge:1957td, Tiomno:1972dq, Sanchez:1976fcl, Futterman:1988ni}, these studies were essential to establish the stability of black holes, understanding the propagation of waves in curved spacetimes, and predicting the characteristic oscillations now known as quasinormal modes which are currently used, for example, in the modelling of black hole mergers \cite{Berti:2005ys}. 
\par In this work, we estimated the radius and energy of the quantum atmospheres of black holes by computing the $S$-matrix \cite{Newton:1982qc, Rakityansky2022} of a massless scalar field in Schwarzschild spacetime. As the $S$-matrix encodes information about the black-hole effects on the scalar field, we can extract its contribution to the evaporation process. We found a new derivation of the Hawking radiation by computing the emission rate through the $S$-matrix, which gives a Bose--Einstein distribution at $T_{H}=\frac{1}{8\pi G M}$, the Hawking temperature. Besides the Hawking radiation, the $S$-matrix also exhibits antibound states, which we identify as the microscopic origin of the quantum atmosphere of black holes. This identification is supported by the result that the antibound spectrum yields $r_{\text{Atm}}\approx2.77r_{s}$, a value for the quantum atmosphere radius that is in good agreement with those obtained through other methods \cite{Giddings:2015uzr, Hod:2016hdd, Dey:2017yez, Dey:2019ugf}. One of the advantages of the present approach is that the Hawking temperature and the properties of the quantum atmosphere emerge directly from the analytic structure of the $S$-matrix computed in the curved spacetime. Hence, our results suggest that the evaluation of scattering amplitudes in the corresponding curved spacetimes may be a valuable alternative method to investigate quantum aspects of black holes.

The paper is organised as follows: in Sec.~\ref{sec:quantum-scattering}, we construct the essential elements of quantum scattering theory in a Schwarzschild spacetime and compute the corresponding $S$-matrix of a massless bosonic field. Sec.~\ref{sec:poles-S-matrix} contains the discussion on the singularities of the $S$-matrix (the poles and the branch cut, related to antibound and scattered states, respectively)  and their physical interpretation. In Sec.~\ref{sec:Hawking-radiation}, we reobtain the Hawking radiation by computing the emission probability from $S-$matrix, where we show that the existence of the antibound states must be properly handled to preserve unitarity. The result of the emission probability is a Bose--Einstein distribution at $T_{H}=\frac{1}{8 \pi G}$. In in Sec.~\ref{sec:quantum-atmosphere}, we discuss why these antibound states can be interpreted as the quantum atmospheres. Using the antibound spectrum, we estimate the radius and energy density of the atmosphere, obtaining a characteristic radius
$r_{\text{Atm}}=4r_{s}\ln2\approx2.77r_{s}$, in good agreement with previous estimates in the literature \cite{Giddings:2015uzr, Hod:2016hdd, Dey:2017yez, Dey:2019ugf}. Finally, in Sec.~\ref{sec:black-hole-mass}, we briefly discuss how our results for the quantum atmosphere could be reinterpreted as a semiclassical remnant of microscopic black-hole mass renormalisation, followed by final remarks in Sec.~\ref{sec:conclusion}. We use $c=\hbar=k_{B}=1$ throughout the rest of the paper.

%%%%%%%%%%%%%%%%%%%%%%%%%%%%%%%%%%%SECTION%%%%%%%%%%%%%%%%%%%%%%%%%%%%%%%%%%%%%%%%%%%%%
\section{Quantum Scattering Theory in the Schwarzschild Spacetime}\label{sec:quantum-scattering}

In this Section, we investigate the quantum scattering of a massless scalar field in a classical Schwarzschild spacetime. We reduce the classical dynamics of the field in such a curved spacetime, imposing a regularity condition on the event horizon in order to construct the adequate quantum $S$-matrix for the desired physical scenario. We focus on the large-$r$ limit to investigate effects that can be detected by far-away observers, which will be considered in further Sections.

The Schwarzschild black hole -- which may be regarded as a highly excited gravitational quantum state \cite{Bekenstein:1972tm, tHooft:1985QuantumStructureBH, Mukhanov:1986me, Bekenstein:1997bt, Hod:1998vk} -- is described in standard spherical coordinates by the line element
\begin{align}\label{eq:Schwarzschild-line-element}
    ds^{2} &= g_{\mu\nu}dx^{\mu}dx^{\nu} =  \\ \nonumber
    &=-\left(1 - \frac{r_{s}}{r} \right) dt^2 + \left(1 - \frac{r_{s}}{r} \right)^{-1} dr^2 + r^2 d\Omega^2,
\end{align}
where $r_{s}=2GM$, the Schwarzschild radius, is an apparent singularity arising due to our choice of a coordinate system \cite{Carroll:2004st}. In fact, the region $r=r_{s}$ is the event horizon of the black hole, the boundary beyond which no causal signal can reach a distant observer. Therefore, the region in which one can directly access physical phenomena is in principle $r>r_s$ .

As we are interested in the effects induced by a black hole on a massless scalar field, we work with the massless Klein--Gordon equation in the Schwarzschild spacetime,
\begin{equation}\label{eq:KleinGordon-KG}
     \bar{\Box}\Phi=0,
\end{equation}
where $\bar{\Box}= g^{\mu\nu}\nabla_{\mu}\nabla_{\nu}$ is the Laplace--Beltrami operator in curved spacetimes and the metric $g_{\mu\nu}$ is defined on \eqref{eq:Schwarzschild-line-element}. 
In this framework, we are considering a quantum field ($\Phi$) in a classical background ($g_{\mu\nu}$). This problem may also be seen as a low-energy effective theory. 
In general, wave equations in curved spacetime are difficult to solve, with no general method for finding solutions useful for practical calculations \cite{Friedlander:2010eqa, Poisson:2011nh}. However, for spacetimes with isometries, their generators can be used to reduce wave equations. 

The isometry group of the Schwarzschild solution is 
\begin{equation}\label{eq:isometry-G}
 \mathrm{G} = SO(3) \times U(1),
\end{equation}
which represents the spacetime invariance under spatial rotations and time translations, respectively. The Casimir operators of G are $i\partial_{t}$, $\mathbf{L}^2$ and $\mathbf{L}_{z}$,
\begin{equation}
    [i\partial_{t} \ ;\mathbf{L}^2]=[i\partial_{t} \ ;\mathbf{L}_{z}]= [\mathbf{L}_{z} \ ;\mathbf{L}^2]=0\,.
\end{equation}
These operators commutes with the wave operator,
\begin{equation}
    [\tilde{\Box} \ ;\mathbf{L}^2]=[\tilde{\Box} \ ;\mathbf{L}_{z}]=[\tilde{\Box} \ ;i\partial_{t}]=0\,,
\end{equation}
hence, we can decompose solutions of the wave equation in terms of eigenfunctions of the Casimir operators,
\begin{equation}
    \Phi(x) = e^{-i \omega t}Y^{m}_{l}(\theta,\phi)\psi(r)\,.
\end{equation}
It is important to emphasise that this is not an arbitrary Ansatz, but the functional form of the eigenmodes of the operator $g^{\mu\nu}\nabla_{\mu}\nabla_{\nu}$, determined by spacetime symmetries.

To implement the methods from quantum scattering theory, it is useful to rewrite the differential equation for the radial modes $\psi(r)$ as a Schrödinger equation through the redefinition
\begin{equation} 
    \psi(r) = \frac{1}{\omega r_{s}}\frac{u(r)}{\sqrt{1-\frac{r_{s}}{r}}}\,,
\end{equation}
where the normalisation $(\omega r_{s})^{-1}$ is a convention from the Jost theory \cite{Newton:1982qc, Rakityansky2022}. It is also useful to use the dimensionless coordinate
\begin{equation}
    x = \frac{r}{r_{s}} -1\,,
\end{equation}
which is zero at the horizon. Then, using the standard spherical harmonics, \eqref{eq:KleinGordon-KG} can be reduced to an ordinary differential equation for $u(x)$, namely
\begin{equation}\label{eq:Schrodingerfinal}
\begin{split}
-\frac{d^{2}u(x)}{dx^{2}}
+\Bigg[
&\kappa^2
+\frac{2\kappa^2-\ell(\ell+1)-\frac12}{x}
+\frac{\kappa^2+\frac14}{x^2}
\\
&+\frac{\ell(\ell+1)+\frac12}{x+1}
+\frac{1}{4(x+1)^2}
\Bigg]u=0\,,
\end{split}
\end{equation}
where $\kappa:=\sqrt{\omega^2 r_{s}^{2}}$. Since our goal is to investigate the macroscopic effects of quantum fluctuations induced by the black hole, we are interested in observables that can be measured far from the event horizon, at
\begin{equation}
r\gg r_{s} \quad \Leftrightarrow \quad \frac{1}{x} \ll 1\,.
\end{equation}
In order $\mathcal{O}(1/x)^2$, \eqref{eq:Schrodingerfinal} reduces to  an exactly solvable Schrödinger equation \cite{Landau:1991wop, Newton:1982qc}, 
\begin{equation}\label{edo}
-\frac{d^{2}u(x)}{dx^{2}}
+ V_\ell(x)
%\right]
u(x)=\kappa^{2} u(x)\,,
\end{equation}
with
\begin{equation}\label{potential}
V_\ell(x) = -\frac{2\kappa^{2}}{x}
- \frac{\kappa^2}{x^2} + \frac{\ell(\ell+1)}{x^{2}}
\end{equation}
being an effective potential where the first two terms are attractive and the last one, $\frac{\ell(\ell+1)}{x^{2}}$, is repulsive. The resulting potential can be repulsive or attractive, depending on the values of $\kappa$ and $\ell$. 

The differential equation \eqref{edo} have two solutions, $M_{\gamma_{l}(\kappa),\,-i\kappa}(2i\lambda x)$ and $W_{\gamma_{l}(\kappa),\,-i\kappa}(2i\lambda x)$, Whittaker functions \cite{whittaker1996course} of the first and second kinds, respectively. The label $\gamma_{l}(k)$ is defined by 
\begin{equation}
\gamma_{\ell}(\kappa)
:=
\sqrt{\left(\ell+\frac12\right)^2-\kappa^2}\,,
\end{equation}

The scattering amplitude is obtained by imposing the boundary conditions that $u(x)$ must be regular at the event horizon, $x=0$. From the mathematical point of view, these boundary conditions are necessary to ensure that \eqref{edo} leads to a well-defined Sturm--Liouville problem, allowing its use to compute the spectrum and also Green functions \cite{Newton:1982qc, Rakityansky2022}. As $W_{\,i\kappa,\gamma_{\ell}(\kappa)}
\!\left(-2i\kappa x\right)$ is singular at the horizon, the radial mode of each $\ell-$wave is simply given by
\begin{equation}
u_{\kappa, \ell}(x) = M_{\,i\kappa,\gamma_{\ell}(\kappa)}
\!\left(-2i\kappa x\right),\,
\end{equation}
More properties of the Whittaker functions are presented in the Appendix \ref{ap1}. 

With the modes in hand, we can employ the Jost decomposition \cite{Newton:1982qc, Rakityansky2022}, where solutions are written in terms of asymptotic states, with which we can straightforwardly compute the $S$-matrix \cite{Newton:1982qc, Rakityansky2022}. In general, the modes are decomposed as 
\begin{equation}
    u_{\kappa, \ell}(x) = F^{\ell}_{+}(k) \varphi_{+}(k;x) + F^{\ell}_{-}(k) \varphi_{-}(k;x)\,,
    \label{jostform}
\end{equation}
with $\varphi_{\pm}(k;x)$ being the so-called outgoing and ingoing Jost solutions, \textit{i.e.} solutions of the wave equation defined by their asymptotic behaviour, while $F_{\pm}(k)$ are called Jost functions, the corresponding expansion coefficients. Then, the Green function is constructed from the Jost solutions, and the $S$-matrix in terms of the Jost functions, as
 \begin{equation}
    S_{\ell}(\kappa) = \frac{F^{\ell}_{-}(\kappa)}{F^{\ell}_{+}(\kappa)}\,.
    \label{JostS}
\end{equation}
A pedagogical example of the Jost method employed on the paradigmatic hydrogen atom is presented in the Appendix \ref{ap2}. Here, we compute the Jost functions and solutions of the scalar field from an identity that connects both Whittaker functions:

\begin{equation}
\begin{aligned}
M_{i\kappa,\gamma_l(k)}&(2i\kappa x) = \Gamma\!\left(1+2\gamma_l(k)\right) \bigg[ \quad \\
&\frac{e^{\pm\left(i\kappa-\gamma_l(k)-\frac12\right)\pi i}}{\Gamma\!\left(\frac12+\gamma_l(k)+i\kappa\right)} \, W_{i\kappa,\gamma_l(k)}(2i\kappa x) \,+ \\[2mm]
&\frac{e^{\pm i\kappa\pi i}}{\Gamma\!\left(\frac12+\gamma_l(k)-i\kappa\right)} W_{-i\kappa,\gamma_l(k)}\!\left(e^{\pm\pi i}2i\kappa x\right)\bigg]\,.
\label{eq:connectionformula}
\end{aligned}
\end{equation}
Then, we obtain respectively the incoming and outgoing Jost solutions,
\begin{equation}
\begin{aligned}
    \varphi_{-}(\kappa;x) &= W_{-i\kappa,\gamma_l(\kappa)}(2i\kappa x) \ \text{and}\\
    \varphi_{+}(\kappa;x) &= W_{i\kappa,\gamma_l(\kappa)}(-2i\kappa x)\,,
\end{aligned}
\end{equation}
and the corresponding Jost functions
\begin{equation}
\begin{aligned}
    F^{\ell}_{+}(\kappa) &= \frac{\Gamma\!\left(1+2\gamma_l(\kappa)\right)}{\Gamma\!\left(\frac12+\gamma_l(\kappa)+i\kappa\right)} e^{\left(i\kappa-\gamma_l(\kappa)-\frac12\right)\pi i} \ \text{and}\\
    F^{\ell}_{-}(\kappa) &= \frac{\Gamma\!\left(1+2\gamma_l(\kappa)\right)}
     {\Gamma\!\left(\frac12+\gamma_l(\kappa)-i\kappa\right)}
e^{i\kappa\pi i}.
\end{aligned}
\end{equation}
Notice that one can easily check that $\varphi_-$ is associated with incoming modes, for example, through its asymptotic expansion,
%which is the incoming solution, as we can see through its asymptotic expansion,

\begin{equation}
     W_{-i\kappa,\,\gamma_{\ell}(\kappa)}(2i\omega r_{s} x) \sim e^{-i\omega r_{s} x}(2i\omega r_{s} x)^{-i\kappa}.
\end{equation}
Finally, from the Jost functions \eqref{eq:Jost-functions}, we construct the $S$-matrix \eqref{JostS}:
\begin{equation}
S_{\ell}(\kappa) = \frac{F_{-}^{\ell}(\kappa)}{F_{+}^{\ell}(\kappa)} = e^{-i\pi (\gamma_{\ell}(k)+\frac{1}{2})}\frac{\Gamma\!\left(\gamma_{\ell}(\kappa)+\frac12+i\kappa\right)}{\Gamma\!\left(\gamma_{\ell}(\kappa)+\frac12-i\kappa\right)}\,,
\end{equation}
where $S_{\ell}(\kappa)$ are the diagonal elements of the $S-$matrix in the basis $\{\kappa;\ell;m\}$. Each of these elements give the transmission and reflection coefficients of each $\ell-$mode and can also be used to compute the partial-wave expansion. Here, we are interested in the emission of thermal radiation by the black hole, which can be seen as a tunneling process, as Parikh and Wilczek showed through the WKB method to derive the Hawking radiation \cite{Parikh:1999mf, Angheben:2005rm, Arzano:2005rs}. 

Although the laws of quantum mechanics and the rotation group constrain the angular momentum to integer or semi-integer values \cite{Landau:1991wop, FaddeevYakubovskii2009}, genuine physical results can be obtained through the analytic continuation of $\ell$ to the complex plane \cite{Newton1962, Newton:1982qc, squires1963complex, Rakityansky2022}. The effectiveness of this method is not due to the existence of eerie particles with complex angular momentum, but rather an artifice for computing physical effects that only appear beyond all orders in the partial-wave expansion. Therefore, such an approach is more adequate if we expect to recover the Hawking radiation as a tunnelling process through the attractive potential in the neighbourhood of the event horizon. 

The $\ell$ dependence of the effective potential (\ref{potential}) comes through the factor $\ell(\ell+1)$, such that this contribution is attractive for $\ell \in (-1,0)$ and repulsive otherwise, with the attraction being maximum for $\ell = \ell^* := -1/2$. In a first approach, it may be expected that a tunneling process is more prominent than scattering states when the potential is maximally attractive. Indeed, the potential $V_{\ell^*}$ is exactly the \textit{approximate potential} that Sanchez prescribed in \cite{Sanchez:1976fcl} to reproduce the correct behaviours both for $r\rightarrow r_S$ and $r\rightarrow +\infty$ and to compute an absorption coefficient with the correct Hawking temperature. Besides recovering the effective potential found by Sanchez \cite{Sanchez:1976fcl}, the value $\ell=-\frac{1}{2}$ is also associated with pair creation. In general, the $N-$particle threshold occurs when $\ell = -\frac{1}{2}\big(3N-5\big)$, as proven by V. Gribov and I. Pomeranchuk \cite{Gribov1962Regge, squires1963complex}. Thus, in our case, the effective potential $V_{\ell=-\frac{1}{2}}(x)$ is related to the creation of a pair of particles by the black hole.

For the reasons given above, we will work with $\ell^*=-\tfrac12$ in what follows. Then, the Jost functions read simply
\begin{equation}\label{eq:jost-l-meio}
\begin{aligned}
    F_{-\frac{1}{2}}^{+}(\kappa) &= \frac{\Gamma\!\left(1+2i\kappa\right)}
     {\Gamma\!\left(\frac12+2i\kappa\right)} e^{-\frac12\pi i} \ \text{and}\\
    F_{-\frac{1}{2}}^{-}(\kappa) &= \frac{\Gamma\!\left(1+2i\kappa\right)\kappa}{\Gamma\!\left(\frac{1}{2}\right)} e^{+ i\kappa\pi}.
\end{aligned}
\end{equation}
It is interesting to remark that, in the limit $\ell=-\frac{1}{2}$, only the outgoing Jost function, $ F_{-\frac{1}{2}}^{+}(\kappa)$, has zeros. As the zeros of the Jost functions are signatures of excitations (such as bound and antibound states, which will be discussed in the next section), this is an indication that the black hole only emits radiation. With these Jost functions, the $S$-matrix becomes
\begin{equation}\label{eq:S-matrix-l=-1/2}
    S_{-\frac{1}{2}}(\kappa) =
    e^{i\frac{\pi}{2}-\pi i\kappa}
    \frac{\Gamma\left(\frac{1}{2}-2i\kappa\right)}
    {\Gamma\left(\frac{1}{2}\right)}\,.
\end{equation}
In the above, we already chose the correct branch on the connection formula between the Whittaker functions, which will be indicated in equation \eqref{eq:norm-S-matrix}.

\section{Singularities of the $S$ Matrix: \\  Antibound States and Black-Hole Evaporation}\label{sec:poles-S-matrix}

The analytic structure of the $S$-matrix encodes information about all types of states, including scattered, resonant, bound, and antibound states \cite{Landau:1991wop, Newton:1982qc}. They are classified by both the type of singularity and its location on the complex plane, cf. Tab.~\ref{tab:states-and-singularities}. We show that the $S$-matrix (\ref{eq:S-matrix-l=-1/2}) exhibits only scattering states -- the Hawking radiation -- and antibound states, which we will argue that are the microscopic constituents of the black hole quantum atmosphere. As a comparison, the hydrogen atom $S$-matrix exhibits bound states and scattering states, as we discuss in Appendix \ref{ap2}.

\begin{table}[h!]
\centering
\begin{tabular}{|c|c|c|}
\hline
\textbf{State} & \textbf{Singularity} & \textbf{Position} \\
\hline
Scattering & Branch Cut & $k > 0,\quad k \in \mathbb{R}$ \\
\hline
Bound State & Pole & $k = i\kappa,\quad \kappa \in \mathbb{R}^{+}$ \\
\hline
Antibound State & Pole & $k = -i\kappa,\quad \kappa  \in \mathbb{R}^{+}$ \\
\hline
Resonance & Pole & $k = k_r - i k_i,\quad k_r, k_i \in \mathbb{R}^{+}$ \\
\hline
Antiresonance & Pole & $k = -k_r - i k_i,\quad k_r, k_i \in \mathbb{R}^{+}$ \\
\hline 
\end{tabular}
\caption{Physical interpretation of the singularities of the $S$-matrix and its position in the complex plane. The anti-bound, resonances, and antiresonances are in the Unphysical region of the complex plane (the second Riemann sheet), as these excitations do not correspond to square integrable wavefunctions \cite{Landau:1991wop, Newton:1982qc}.}
\label{tab:states-and-singularities}
\end{table}
\par Antibound excitations were discovered by H. Bethe and Peiels in their investigation on the slow energy neutron-proton ($n-p$) scattering \cite{BethePeierls1935, Bethe:1949yr}. They correspond to excitations in the vicinity of becoming a bound state (scattered state), if the attraction is increased (decreased). The low-energy $n$-$p$ system \cite{BethePeierls1935} provides a classic example of this phenomenon. Depending on the spin-isospin channel, the interaction may support either a bound or an antibound state. In the triplet channel, the nuclear attraction is sufficiently strong to form the deuteron, the only bound state of the two-nucleon system. In the singlet channel, however, the attraction is slightly weaker and no bound state is formed. Nevertheless, the interaction remains strong enough to generate an antibound state very close to the scattering threshold. This proximity to threshold is responsible for the unusually large scattering length observed in low-energy $n$-$p$ scattering and provides one of the clearest physical manifestations of an antibound state. Coming back to the problem at hand, we argue that this mechanism is a natural candidate for a description of the black-hole quantum atmosphere. Since antibound states enhance the scattering cross-section — and, consequently, the effective radius of the interaction — they provide an adequate mechanism to generate an extended atmosphere around the black hole. In the next sections, we will show that the antibound spectrum gives a result for the atmosphere radius with good agreement with other values found via different methods \cite{Giddings:2015uzr, Dey:2017yez, Dey:2019ugf}.

\par With the semiclassical $S$-matrix found in \eqref{eq:S-matrix-l=-1/2}, we now discuss the analytic structure of a scalar field around the Schwarzschild black hole. The branch point associated with scattering states arises from $\kappa:=\sqrt{\omega^2 r_{s}^{2}}$ while the poles associated with the discrete spectrum of excitations originate from the $\Gamma(z)$ function. As $\Gamma(z)$ has poles for $z= -n, \ n \in \mathbb{N}$,  the poles of the $S$-matrix are
\begin{equation}
    \omega_{n} = - \frac{i}{2r_{s}}\Big( n+\frac{1}{2}\Big), \ n \in \mathbb{N}\,.
\end{equation}
They are all equally spaced purely imaginary poles with negative imaginary part, therefore, they describe antibound excitations. They are not states in the strict sense because they are associated to non-square-integrable wavefunctions \cite{Newton:1982qc}, but are still physically relevant, as we just recalled from the hydrogen atom, and are usually related to unstable processes.
\begin{figure}[ht]
    \centering
    \includegraphics[width=1\linewidth]{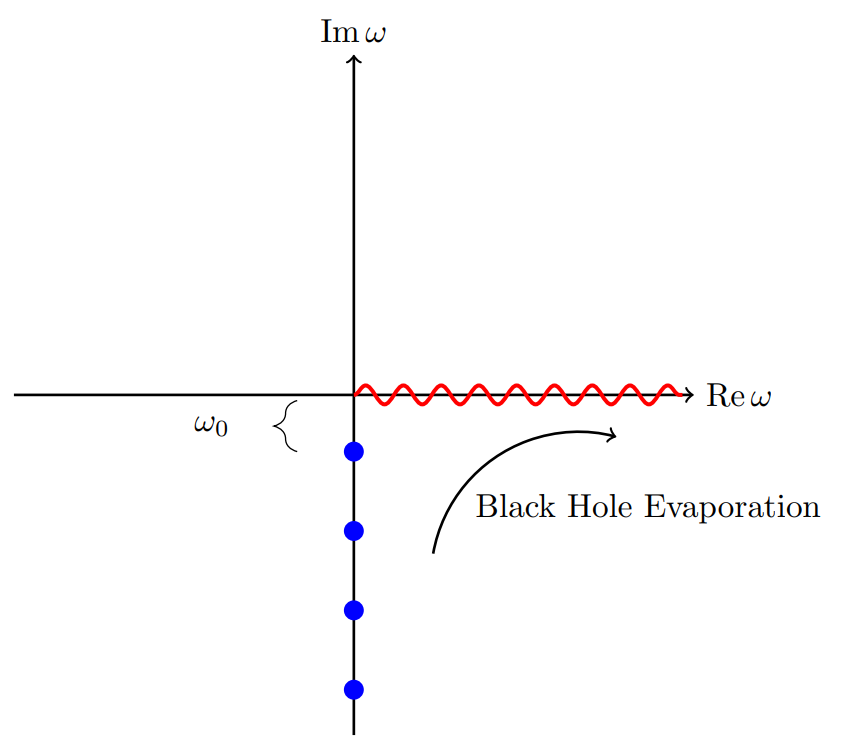}
    \caption{The blue dots denote the antibound poles located on the negative imaginary axis, while the red curve represents the branch cut associated with the scattering states. During black hole evaporation, the first antibound state, $\omega_{0}=-i/(4r_{s})$, is the nearest excitation to the physical sheet and therefore the first to transition into a scattering state. The energy gap associated with this transition is used to recover the Bose--Einstein distribution from the $S$-matrix.}
    \label{bhevap}
\end{figure}
We discussed that antibound states can in principle become either bound states or scattering states. However, the $S$-matrix in this case only contains scattering and antibound states. Indeed, the black-hole evaporation itself is a suitable mechanism for antibound states to become scattering instead of bound. As the black hole evaporates emitting an energy $\epsilon$, it loses mass
\begin{equation}
M \rightarrow M-\epsilon,
\end{equation}
which implies a decrease of the Schwarzschild radius,
\begin{equation}
r_{s}=2GM \rightarrow 2G(M-\epsilon).
\end{equation}
As a consequence, the effective potential becomes less attractive. A reduction of the attraction drives the antibound states towards the scattering continuum, therefore, black-hole evaporation naturally induces transitions from antibound to scattering states. 

We have shown that there is a natural physical mechanism for Hawking radiation and quantum atmospheres to find their microscopic origin on antibound states of fields near a black hole becoming scattering. In what follows, we will back this claim up with explicit computations.

%%%%%%%%%%%%%%%%%%%%%%%%%%%%%%%%%SECTION%%%%%%%%%%%%%%%%%%%%%%%%%%%%%%%%%%%%%%%
\section{Evaluation of the Hawking Radiation Emission Rate Using the $S$-Matrix}\label{sec:Hawking-radiation}

\par The usual approaches to derive Hawking radiation in Schwarzschild spacetimes involve either Bogoliubov coefficients of mode expansions \cite{Birrell:1982ix, Wald:1995yp} or path integral of a scalar field in the quantum field theory in curved spacetimes formalism \cite{Hartle:1976tp, Witten:2024upt}, or, alternatively, a semiclassical approach using WKB \cite{Parikh:1999mf, Angheben:2005rm, Arzano:2005rs}. In this Section, we show a novel derivation of the Hawking radiation through the $S$-matrix formalism.

Given any physical system, the $S-$matrix is defined as the operator $\hat{S}$ that connects initial and final states, $\ket{\mathrm{In}}$ and $\ket{\mathrm{Out}}$:
\begin{equation} 
    \hat{S}\ket{\text{In}} = \ket{\text{Out}}\,.
\end{equation}
Taking the square of the norm on both sides for a particular pair of initial and final states with energy $\omega$, we obtain the probability of emission of a quanta with this energy (also called Gamow factor or Sommerfeld enhancement factor, depending of the context),
\begin{equation}  
    P(\omega) =  \norm{\mathbf{S}} =\frac{\norm{\ket{\text{Out},\omega}}^{2}}{\norm{\ket{\text{In},\omega}}^{2}}.
\end{equation}

In this vein, we now obtain the emission rate of the Hawking radiation from the semiclassical $S$-matrix \eqref{eq:S-matrix-l=-1/2}. For this, one needs to take into account the presence of excitations in the unphysical sheet, which can lead to problems with unitarity. A direct calculation of the modulus of the $S$-matrix  gives
\begin{equation}\label{eq:norm-S-matrix}
|S(\omega)|^{2}
=
\frac{e^{-2\pi\kappa}}{\pi}
\left|
\Gamma\!\left(\frac{1}{2}-2i\kappa\right)
\right|^{2} = \frac{e^{-4\pi r_{s}\omega}}{1+e^{-4\pi r_{s}\omega}}
\end{equation}
where we used the Gamma function reflection identity \cite{NIST:DLMF}
\begin{equation}
    |\Gamma\Big(\frac{1}{2}-iy\Big)\Big|^{2} = \frac{\pi}{\cosh{\pi y}}\,.
\end{equation}
If we chose the other branch for the Whittaker functions connection formula, as discussed around \eqref{eq:jost-l-meio}, the sign of the real exponential in \eqref{eq:norm-S-matrix} would be positive and the system would be unphysical \cite{Landau:1991wop, Newton:1982qc, Rakityansky2022}. This gives us the correct branch choice.

The emission rate \eqref{eq:norm-S-matrix} gives a Fermi--Dirac distribution with temperature
\begin{equation}
T_H=\frac{1}{4\pi r_{s}}=\frac{1}{8\pi GM}\,,
\end{equation}
the Hawking temperature.  This was obtained solely from the scattering amplitude analysis. To the best of our knowledge, this provides a novel derivation of the Hawking temperature, showing that the thermal scale of black-hole evaporation is already encoded in the analytic structure of the $S$-matrix of a scalar field around the black hole. However, notice that \eqref{eq:norm-S-matrix} exhibits a Fermi--Dirac statistics rather than the Bose--Einstein one expected for a massless scalar field. This is not unique to our approach, but also arises in the pioneering work of Norma Sanchez \cite{Sanchez:1976fcl}. The origin of this discrepancy can be traced to the poles of the $S$-matrix lying on the second Riemann sheet, where states are non-normalisable and the standard notion of unitarity is not respected. Therefore, it may be expected that it is possible to obtain the correct thermal scaling, but with wrong statistics, violating the spin-statistics theorem \cite{Schwartz:2014sze}, which relies on unitarity and causality assumptions. In such cases, one has to take the antibound state into the physical sheet. Here, we simply shift the spectrum upwards by the value of the first gap, 

in this case, $\omega\rightarrow\omega+\frac{i}{4r_s}$. Then, the first antibound state will be on the verge of becoming a physical state, either scattering or bound, depending on the system. The shifted $S$-matrix modulus reads
\begin{equation}
    \Big|S\Big(\omega+\frac{i}{4 r_{s}}\Big)\Big|^{2} =e^{-2\omega \pi r_{s}} \frac{\Big| \Gamma\big(2i\omega r_{s}\big)\Big|^{2}}{\pi}\,.
\end{equation}
Using another Gamma function reflection identity \cite{NIST:DLMF},
\begin{equation}
    |\Gamma\Big(-iy\Big)\Big|^{2} = \frac{\pi}{y\sinh{\pi y}}\,,
\end{equation}
we obtain
\begin{equation}
    P(\omega) = \frac{1}{\omega r_{s}}\frac{e^{-4\pi r_{s}\omega}}{1-e^{-4\pi r_{s}\omega}}\,.
\end{equation}
Therefore, shifting the $S$-matrix by the gap of the first antibound state transformed the distribution from a Fermi--Dirac into a Bose--Einstein distribution with temperature $T_{H}=\frac{1}{8\pi G M}$, which is now compatible with the spin-statistics theorem for a bosonic field. This finalises our new derivation of the Hawking radiation spectrum only from the $S$-matrix of a scalar field on a Schwarzschild spacetime. Finally, notice that the emission diverges in the IR limit, $\omega \to 0$. However, this is not a physical divergence since the energy of the Hawking radiation is bounded from below, $\epsilon > T_{H}$, providing a natural IR regularisation.

\section{ Radius of the Quantum Atmosphere of Black Holes and the Antibound States}\label{sec:quantum-atmosphere}

The nature of the quantum atmospheres of black holes is of great interest for a deeper understanding of astrophysical black holes \cite{Giddings:2015uzr,Hod:2016hdd,Dey:2017yez,Dey:2019ugf} and the Information Puzzle \cite{Giddings:2015uzr, Zhang:2026ixo}. The central idea is that the surface of the quantum atmosphere is the actual source of the Hawking radiation, rather than the event horizon itself. The goal of this section is to estimate the radius of the black-hole quantum atmosphere using the antibound spectrum obtained in the previous sections.

Given the prediction of the quantum atmospheres of black holes, the natural question is the size of this atmosphere. Different estimations for the value of the radius of the quantum atmosphere of a Schwarzschild black hole were found using a plethora of methods, 

\begin{table}[ht]
\centering
\renewcommand{\arraystretch}{1.2}
\begin{tabular}{|c|c|}
\hline
\textbf{$r_\text{Atm}$} & \textbf{Method} \\
\hline
$2.67\,r_s$ &
Stefan--Boltzmann law and numerical analysis \cite{Giddings:2015uzr} \\
\hline
$2.19\,r_s$ &
Tidal-force analysis \cite{Dey:2017yez,Dey:2019ugf} \\
\hline
$2.18\,r_s$ &
Canonical quantisation (Hartle--Hawking vacuum) \cite{Dey:2017yez} \\
\hline
$2.16\,r_s$ &
Canonical quantisation (Unruh vacuum) \cite{Dey:2017yez} \\
\hline
\end{tabular}
\caption{Estimates of the quantum atmosphere radius for a Schwarzschild black hole obtained using different theoretical methods. The quoted values correspond to massless scalar fields; the atmosphere radius depends on the spin of the emitted field \cite{Gingrich:2023qae}.}
\label{ramt}
\end{table}

The values in Table \ref{ramt} indicate that the radius of the quantum atmosphere is a few times larger than the Schwarzschild radius, suggesting that the Hawking radiation is not generated strictly at the event horizon, but rather in an extended region surrounding the black hole. Therefore, any microscopic description of the quantum atmosphere should naturally involve length scales larger than $r_{s}$, while remaining in the same order of magnitude. As we shall see, the antibound spectrum provides a natural mechanism to generate such a scale.

We argued in previous Sections that the antibound states can be regarded as microscopic constituents of the quantum atmosphere, leading to a reproduction of the Hawking temperature and of the Hawking radiation as the antibound become scattering states. In this picture, the first antibound state is the nearest excitation to the physical sheet of the $S$-matrix and therefore the closest to becoming a real scattering state. On the other hand, antibound states with large $n$ are located deeper in the unphysical region of the complex-frequency plane and are farther from the threshold for transition. Hence, while the first antibound states controls the onset of the evaporation process, at any given moment, all antibound states contribute to the bulk of the quantum atmosphere. With this interpretation in mind, we can use the antibound spectrum to estimate the radius of the quantum atmosphere.

A first and crude quantum-atmosphere radius estimative can be made using the first antibound state, $\omega_{o}=-\frac{i}{4r_{s}}$, leading to
 \begin{equation}
     \Bigg|\frac{1}{\omega_{o}}\Bigg| = 4r_{s},
 \end{equation}
which is in the same length scale as the predictions of Giddings \cite{Giddings:2015uzr} and Rey \cite{Dey:2017yez}. As we just argued, a better estimative takes into account that all antibound states constitute the quantum atmosphere. Summing over all contributions, we define
\begin{equation}
    r_{\text{Atm}} := \sum_{n=0}^{\infty}\frac{1}{\omega_{n}} = 2r_{s} \sum_{n=0}^{\infty}\frac{1}{n+\frac{1}{2}}\,.
\end{equation}
Comparing with the harmonic series, it is easy to see that the sum above diverges logarithmically \cite{Schwartz:2014sze}. However, we can still find a finite value for the atmosphere radius from zeta function regularisation \cite{Hawking:1976ja}. This procedure is implemented by first defining the analytic continuation of the sum,
\begin{equation}
    \sum_{n=0}^{\infty}\frac{1}{(n+\frac{1}{2})^{s}} = \zeta\Big(s,\frac{1}{2}\Big)\,,
\end{equation}
and its relation to $r_{\text{Atm}}$,
\begin{equation}
    r_{\text{Atm}}= 2r_{s} \lim_{s\to 1^{+}}\zeta\Big(s;\frac{1}{2}\Big)\,.
\end{equation}
To compute $r_{\text{Atm}}$, one must simply evaluate the expansion of $\zeta_{H}(s,\frac{1}{2})$ around $1^{+}$. Thus, we obtain, for the radius of the quantum atmosphere,
\begin{equation}\label{eq:QA-radius}
\begin{aligned}
    r_{\text{Atm}}= 4r_{s}\ln2 &+ \text{Divergent Part}=\\
    2.77259r_{s} &+ \text{Divergent Part}\,.
\end{aligned}
\end{equation}
This result is within approximately $5\%$ of the value predicted by Giddings \cite{Giddings:2015uzr} and also in good agreement with the values presented in the Table \ref{ramt}.

%%%%%%%%%%%%%%%%%%%%%%%%%%%%%%%%%%%%%%%%%%%%%%%%%%%%%%%%%%%%%%%%%%%%%%%%%%%%

\section{Semiclassical correction to the Black Hole Mass}\label{sec:black-hole-mass}

The quantum atmosphere of the black hole -- the volume between $r_{\text{Atm}}$ and $r_{s}$ -- is the actual source of the Hawking radiation. Therefore, it must contribute to the energy of the system black hole and its atmosphere. 

The energy of the quantum atmosphere can be estimated from the antibound states by regularising the sum of the $|\omega_{n}|$,
\begin{equation}
\textit{E}_{\text{Atm}}=
\sum_{n=0}^{\infty} |\omega_n|
\;\;=
\frac{1}{2r_s}\zeta\!\left(-1,\frac12\right)\,.
\end{equation}
The function $\zeta\!\left(-1,\frac12\right)$ have a finite value, leading to, 
\begin{equation}
E_{\text{Atm}}= \frac{1}{48r_{s}}\,,
\end{equation}
the energy of the black-hole quantum atmosphere. To compute the energy density, we must use the radius of the quantum atmosphere \eqref{eq:QA-radius} to compute its volume,
\begin{equation}
    V_{\text{Atm}}= 4\pi \int_{r_{s}}^{r_{\text{Atm}}}\frac{dr \ r^{2}}{\sqrt{1-\frac{r_{s}}{r}}} \approx 30 V_{\text{BH}}\,.
\end{equation}
Then,
\begin{equation}
    u_{\mathrm{Atm}}
    =
    \frac{E_{\text{Atm}}}{V_{\text{Atm}}}
    \approx
    \frac{1}
    { r_s^{4}}
    \propto
   + T_H^{4}\,,\label{eq:uatm}
\end{equation}
which shows that the energy density of the quantum atmosphere is proportional to $T_{H}^{4}$ and positive, as any thermal gas of massless bosons. This result is remarkable because it emerges directly from the antibound spectrum of the black hole. Therefore, although the quantum atmosphere is described microscopically in terms of antibound excitations, its macroscopic behaviour is indistinguishable from that of ordinary thermalised matter, as suggested by Biggs and Maldacena \cite{biggs}. This provides further evidence that the antibound states are the microscopic degrees of freedom responsible for the thermodynamic properties of the quantum atmosphere and, consequently, for the Hawking radiation itself.

It is well known that Schwarzschild black holes possess a negative heat capacity \cite{Witten:2024upt}, a hallmark of a thermodynamically unstable system. The quantum atmosphere, however, behaves differently. As equation ~(\ref{eq:uatm}) shows, its energy density scales as $u_{\text{Atm}}\propto +T_{H}^{4}$, leading to a positive heat capacity, as expected for a thermal gas of massless bosons. Therefore, while the black hole becomes hotter as it loses energy, the quantum atmosphere behaves as ordinary thermalised matter. This distinction further supports the picture in which the quantum atmosphere acts as an intermediate thermal region between the black hole and the asymptotic observer.

As we showed in the previous section, the computation of $r_{A}$ leads to a logarithmic divergence. From the perspective of quantum field theory, the radius of the quantum atmosphere can be regarded as a renormalised value of the event horizon radius, $r_{s}$. Since $r_{s}=2GM$ and the logarithmic divergence does not appear in the renormalisation of the black-hole mass, it is natural to associate this divergence with the renormalisation of the gravitational coupling constant $G$. Therefore, the quantum atmosphere can be interpreted as a {semiclassical manifestation of the quantum corrections to the classical Schwarzschild geometry. It is worth mentioning that perturbative calculations of quantum gravity also lead to logarithmic divergences in the renormalisation of the gravitational coupling \cite{Goroff:1985th,Percacci:2017fkn}. Although our analysis does not constitute a derivation of the running of $G$, the appearance of the same type of divergence suggests that quantum atmospheres may encode information about the renormalisation of gravity on black-hole spacetimes}.

%%%%%%%%%%%%%%%%%%%%%%%%%%%%%%%%%%%SECTION%%%%%%%%%%%%%%%%%%%%%%%%%%%%%%
\section{Conclusions and Final Remarks}\label{sec:conclusion}

\par In this paper, we analysed the thermodynamics of Schwarzschild black holes by computing the $S$-matrix of a massless bosonic field in the curved spacetime. From the analytic structure of the scattering amplitude, we obtained the Hawking temperature and the emission rate of Hawking radiation. This result provides a new derivation of black-hole evaporation based entirely on scattering theory. Unlike the traditional approaches based on Bogoliubov transformations \cite{Hawking:1975vcx},  path integral quantisation \cite{Hartle:1976tp, Witten:2024upt} or on the WKB quantisation \cite{Parikh:1999mf, Arzano:2005rs, Jiang:2005ba}, the present formalism extracts the thermodynamic properties of the black hole directly from the singularities of the $S$-matrix. In this sense, our results establish a direct connection between black-hole thermodynamics and quantum scattering theory.

Beyond reproducing Hawking radiation, we showed that the same scattering framework predicts the existence of a quantum atmosphere surrounding the black hole. Within our interpretation, the antibound states play the role of the microscopic constituents of this atmosphere. Using their spectrum, we estimated the radius of the quantum atmosphere, obtaining
\begin{equation}
r_{\text{Atm}}=4r_{s}\ln 2 \approx 2.77\,r_{s},
\end{equation}
which is in good agreement with previous estimates in the literature \cite{Giddings:2015uzr, Hod:2016hdd, Dey:2017yez, Dey:2019ugf, Gingrich:2023qae} computed using other methods. Remarkably, although the microscopic description is given in terms of antibound excitations, we find that the atmosphere behaves macroscopically as a thermal gas of massless bosons, which is in agreement with \cite{biggs} . Therefore, the thermodynamic properties of the atmosphere emerge naturally from the scattering amplitude of the wave equation in the Schwarzschild spacetime.

The methods developed in this work can be extended to other black holes \cite{Carroll:2004st}. In particular, the Reissner--Nordström case, where spherical symmetry is preserved, and many of the techniques employed here remain applicable. One of the advantages of the present approach is its simplicity: the thermodynamic properties of the black hole are extracted directly from the solutions of a wave equation and the associated $S$-matrix, avoiding the more elaborate machinery usually employed in the study of Hawking radiation. The Kerr geometry presents a greater challenge, as the loss of spherical symmetry makes the treatment of modes on the Kerr spacetime significantly more involved.% 

Finally, the scattering approach developed in this work can also be viewed from the perspective of quantum gravity. In the present analysis, the spacetime geometry is treated classically and the quantum fluctuations of the gravitational field are neglected. Therefore, the results obtained here may be interpreted as the low-energy sector of a more fundamental quantum gravitational description. From this point of view, Hawking radiation emerges as a consequence of the scattering of quantum fields by a curved background. A natural extension of the present work would be to include gravitons in the analysis. Such a generalisation would go beyond the semiclassical approximation and could provide a direct connection between the black-hole evaporation and the microscopic degrees of freedom of quantum gravity.

\begin{acknowledgments}

The author, V. H. Alencar, would like to thank Prof. Pavel Petrov for enlightening discussions on semiclassical methods and scattering theory. The authors would like to express their gratitude to João G. A. Caribé for useful discussions on black-hole vacua. This work was partly funded by the Coordena\c{c}\~{a}o de Aperfei\c{c}oamento de Pessoal de N\'ivel Superior - Brasil (CAPES) - Finance Code 001. G.P. is supported by the Conselho Nacional de Desenvolvimento Científico e Tecnológico (CNPq) under the grant no.~446877/2024-7. C.A.D.Z. is partially supported by Conselho Nacional de Desenvolvimento Cient\'ifico e Tecnol\'ogico (CNPq) under the grant no.~305610/2025-2. C.A.D.Z. is also funded by Funda\c{c}\~{a}o Carlos Chagas Filho de Amparo \`a Pesquisa do Estado do Rio de Janeiro (Faperj) under Grant no.~E-26/201{.}447/2021 (Programa Jovem Cientista do Nosso Estado).
\end{acknowledgments}

\appendix
\section{Some Properties of Whittaker Functions}
\label{ap1}
 The so-called Whittaker functions, $M_{\kappa,\mu}$ and $W_{\kappa,\mu}$, are the two linearly independent solutions of the linear ordinary differential equation \cite{whittaker1996course,beals2010special,NIST:DLMF},
\begin{equation}\label{eq:Whittaker}
\dv[2]{y(z)}{z}
+
\left(
-\frac{1}{4}
+\frac{\kappa}{z}
+\frac{\tfrac{1}{4}-\mu^2}{z^2}
\right)
y(z)
=
0\,,
\end{equation}
with general solution
\begin{equation}
y(z)
=
c_1\, M_{\kappa,\mu}(z)
+
c_2\, W_{\kappa,\mu}(z),
\end{equation}
where $c_1$ and $c_2$ are constants. Both Whittaker functions are related to  the confluent hypergeometric functions\cite{NIST:DLMF,abramowitz1965handbook},
\begin{equation}\label{eqMF}
M_{\kappa,\mu}(z)
=
z^{\mu+\tfrac{1}{2}} e^{-z/2}
\, {}_1F_1\!\left(
\mu-\kappa+\tfrac{1}{2},\,
2\mu+1;\,
z
\right)\,,
\end{equation}
where ${}_1F_1\!\left(
\mu-\kappa+\tfrac{1}{2},\,
2\mu+1;\,
z
\right)$ is the confluent hypergeometric function of the first kind, and
\begin{equation}
W_{\kappa,\mu}(z)
=
z^{\mu+\tfrac{1}{2}} e^{-z/2}
\, U\!\left(
\mu-\kappa+\tfrac{1}{2},\,
2\mu+1;\,
z
\right)\,,
\end{equation}
where $U\!\left(
\mu-\kappa+\tfrac{1}{2},\,
2\mu+1;\,
z
\right)$ is the confluent hypergeometric function of the first kind. 

The behaviour of the Whittaker function near the origin ($z\to 0$) are given by
\begin{equation}
\begin{aligned}
    M_{\kappa,\mu}(z)&\sim z^{\mu+\frac{1}{2}} \\
    W_{\kappa,\mu}(z)&\sim \frac{\Gamma(2\mu)}{\Gamma\left(\frac{1}{2}+\mu-\kappa\right)}z^{\frac{1}{2}-\mu}.
\end{aligned}
\end{equation}
Hence  $M_{\kappa,\mu}(z)$ is regular and $W_{\kappa,\mu}(z)$ is singular at $z=0$. Near infinity, they behave as
\begin{align}
    M_{\kappa,\mu}(z)&\sim \frac{\Gamma\left(1+2\mu\right)}{\Gamma\left(\frac{1}{2}+\mu-\kappa\right)}e^{\frac{z}{2}}z^{\kappa}\;\;\\
    W_{\kappa,\mu}(z)&\sim e^{-\frac{z}{2}}z^{\kappa}\,.
\end{align}
For $z\to\infty$, $W_{\kappa,\mu}(z)$ is regular and $ M_{\kappa,\mu}(z)$ is singular. 

Moreover, the following connection formula will be useful for our purposes:
\begin{equation}
\begin{aligned}\label{eq:connectionformula}
M_{\kappa,\mu}(z)& = \frac{\Gamma\left(1+2\mu\right)}{\Gamma\left(\frac{1}{2}+\mu+\kappa\right)}e^{\pm\left(\kappa-\mu-\frac{1}{2}\right)\pi i}W_{\kappa,\mu}(z)\nonumber \\
&+\frac{\Gamma\left(1+2\mu\right)}{\Gamma\left(\frac{1}{2}+\mu-\kappa\right)}e^{\pm\kappa\pi i}W_{-\kappa,\mu}(e^{\pm\pi i}z).
\end{aligned}
\end{equation}

\section{$S-$Matrix of the Non-Relativistic Hydrogen Atom} \label{ap2}

In general presentations of quantum mechanics textbooks, the hydrogen spectrum is found by solving the eigenvalue problem given by the corresponding Hamiltonian \cite{Landau:1991wop, FaddeevYakubovskii2009,tong2025quantum}. In order to illustrate the method for the main development of this paper, it is instructive to employ the Jost method to compute the spectrum of the hydrogen atom through its $S$-matrix.  

As the Coulomb field is spherically symmetric and time-independent, the angular and temporal dependences of the eigenfunctions are given by $e^{i\omega t}$ and $Y_{\ell}^{m}(\theta,\varphi)$, reducing the problem to the ODE,
\begin{equation}
   \Bigg( \dv[2]{r}+\frac{2}{r}\dv{r}+k^{2}-\frac{l(l+1)}{r^2}+\frac{2\gamma k}{r} \Bigg)\psi_{k,l}(r)=0\,,
\end{equation}
where $k^{2}=2mE$ and $\gamma=\frac{me^{2}}{|k|}$. This equation can be expressed in the Whittaker form by using the following transformations:
\begin{equation}
    \psi_{\text{k,l}}(r) = \frac{u_{k,l}(r)}{r} \text{ and }    \rho=2i|k|r\,,
\end{equation}
yelding the equation
\begin{equation}\label{eq:whittakerhydrogen}
    \dv[2]{u_{kl}(\rho)}{\rho}+ \Bigg(\frac{\frac{1}{4}-(l+\frac{1}{2})^{2}}{\rho^{2}}-\frac{i\gamma}{\rho}-\frac{1}{4} \Bigg)u_{kl}(\rho)=0\,.
\end{equation}

Comparing with Eq. \eqref{eq:Whittaker} and imposing that $u_{kl}(\rho)$ must be regular at $\rho=0$, one finds the regular solution nearby $\rho=0$\,,
\begin{equation}
u_{k,\ell}(\rho)=\, M_{-i\gamma,l+\frac{1}{2}}(\rho)\,.
\end{equation}

\par The well-known hydrogen wave functions in terms of associated Laguerre polynomials \cite{Landau:1991wop, tong2025quantum} can be recovered by using the relation between the Whittaker function and the hypergeometric confluent function (\ref{eqMF}),
\begin{equation}
{}_1F_1(-n,\alpha+1,x)
=
\frac{n!\,\Gamma(\alpha+1)}
{\Gamma(n+\alpha+1)}
L_n^{(\alpha)}(x)\,.
\end{equation}

To identity the Jost functions and solutions and evaluate the $S$-matrix using the solution  \eqref{eq:whittakerhydrogen}, we must use the identity \eqref{eq:connectionformula},
\begin{equation}
\begin{aligned}\label{eq:connectionformula}
M_{i\gamma,l+\frac{1}{2}}(2i|k|r)
&=
\frac{\Gamma\!\left[2(l+1)\right]}{\Gamma\!\left(l+1-i\gamma\right)}
e^{\pi\gamma}e^{-(l+1)\pi i}
W_{-i\gamma,l+\frac{1}{2}}(2i|k|r)\\
&\quad+
\frac{\Gamma\!\left[2(l+1)\right]}{\Gamma\!\left(l+1-i\gamma\right)}
e^{\pi\gamma}
W_{i\gamma,l+\frac{1}{2}}(-2i|k|r)\,.
\end{aligned}
\end{equation}
Using the scattering amplitude rewritten in this way, we can use the Jost form \eqref{jostform} to identify the ingoing and outgoing Jost solutions, respectively $\phi_{-}(k,r)=W_{-i\gamma,l+\frac{1}{2}}(2ikr)$ and $\phi_{+}(k,r)=W_{i\gamma,l+\frac{1}{2}}(-2ikr)$. The Jost functions, $F_{+}(k)$ e $F_{-}(k)$ are then the coefficients of the equation above \cite{Newton:1982qc}:
\begin{equation}
\begin{aligned}
    M_{-i\gamma,l+\frac{1}{2}}(2i|k|r)&=F_{+}(k)\phi_{+}(k,r)+F_{-}(k)\phi_{-}(k,r)\,;\\
    F_{-}(k)&=\frac{\Gamma\left[2(l+1)\right]}{\Gamma\left(l+1-i\gamma\right)}e^{\pi\gamma}e^{-(l+1)\pi i}\\
    F_{+}(k)&=\frac{\Gamma\left[2(l+1)\right]}{\Gamma\left(l+1+i\gamma\right)}e^{\pi\gamma}\,.
\end{aligned}
\end{equation}
Now the $S-$matrix can be computed as
\begin{equation} \label{Sh}
    S_{l}(k)=\frac{ F_{-}(-k)}{F_{+}(k)}=e^{-(l+1)\pi i}\frac{\Gamma\left(l+1+i\gamma\right)}{\Gamma\left(l+1-i\gamma\right)}\,.
\end{equation}

Finally, we can find the spectra of the hydrogen atom by computing the poles of the $S-$matrix. As the poles of the Gamma function are located at the non-positive integers,  the poles of the $S-$matrix are
\begin{equation}
    l+1+i\gamma=-N\,,
\end{equation}
where $N \in \mathbb{N}$. Hence, using the definition of $\gamma$, the poles of the $S-$ matrix are located at
\begin{equation}
    k_{n}=-i\frac{me^{2}}{N+l+1}=-i\frac{me^{2}}{n}\,,
\end{equation}
where we have defined $n:=N+l+1$ ($n \in \mathbb{N}^{*}$) as the principal quantum number. Using the dispersion relation, $E(k)$, we find the spectrum of the nonrelativistic hydrogen atom,
\begin{equation}
    E_{n}=\frac{k^2}{2m}=-\frac{me^{4}}{2m}\frac{1}{n^{2}} \approx - \frac{13.6}{n^{2}}\ \text{eV}\,,
\end{equation}
as expected.

Finally, we compute the emission rate of the unperturbed hydrogen atom using the $S$-matrix, (\ref{Sh}),
\begin{equation}
     |S_{\ell}(k)|^{2}=\Big |\frac{\Gamma\left(l+1+i\gamma\right)}{\Gamma\left(l+1-i\gamma\right)}\Big|^{2}\,.
\end{equation}
Using the reflection identities and the property $\Gamma(x+iy)= \overline{\Gamma(x-iy)}$, which implies in  $|\Gamma\left(l+1+i\gamma\right)|^{2}= |\Gamma\left(l+1-i\gamma\right)|$, we obtain
\begin{equation}
     |S_{\ell}(k)|^{2}=1, \ \forall  \ l  \ ,
\end{equation}

\appendix

%\nocite{*} Queremos colocar somente o que está no arquivo

\bibliographystyle{apsrev4-2}
\bibliography{ref}% Produces the bibliography via BibTeX.

@PREAMBLE{
 "\providecommand{\noopsort}[1]{}" 
 # "\providecommand{\singleletter}[1]{#1}%" 
}

@article{tHooft:1985QuantumStructureBH,
    author = "t Hooft, Gerard",
    title = {{On the quantum structure of a black hole}},
    journal = "Nucl. Phys. B",
    volume = "256",
    pages = "727--745",
    year = "1985",
    doi = "10.1016/0550-3213(85)90418-3",
    publisher = "North-Holland Publishing Company"
}

@article{Arzano:2005rs,
    author = "Arzano, Michele and Medved, A. J. M. and Vagenas, Elias C.",
    title = "{Hawking radiation as tunneling through the quantum horizon}",
    eprint = "hep-th/0505266",
    archivePrefix = "arXiv",
    doi = "10.1088/1126-6708/2005/09/037",
    journal = "JHEP",
    volume = "09",
    pages = "037",
    year = "2005"
}

@article{Jiang:2005ba,
    author = "Jiang, Qing-Quan and Wu, Shuang-Qing and Cai, Xu",
    title = "{Hawking radiation as tunneling from the Kerr and Kerr-Newman black holes}",
    eprint = "hep-th/0512351",
    archivePrefix = "arXiv",
    doi = "10.1103/PhysRevD.73.064003",
    journal = "Phys. Rev. D",
    volume = "73",
    pages = "064003",
    year = "2006",
    note = "[Erratum: Phys.Rev.D 73, 069902 (2006)]"
}

@article{Hawking:1975vcx,
    author = "Hawking, S. W.",
    editor = "Gibbons, G. W. and Hawking, S. W.",
    title = "{Particle Creation by Black Holes}",
    doi = "10.1007/BF02345020",
    journal = "Commun. Math. Phys.",
    volume = "43",
    pages = "199--220",
    year = "1975",
    note = "[Erratum: Commun.Math.Phys. 46, 206 (1976)]"
}

@article{Hawking:1976ja,
    author = "Hawking, S. W.",
    title = "{Zeta Function Regularization of Path Integrals in Curved Space-Time}",
    reportNumber = "PRINT-77-0293 (CAMBRIDGE)",
    doi = "10.1007/BF01626516",
    journal = "Commun. Math. Phys.",
    volume = "55",
    pages = "133",
    year = "1977"
}

@article{Hod:1998vk,
    author = "Hod, Shahar",
    title = "{Bohr's correspondence principle and the area spectrum of quantum black holes}",
    eprint = "gr-qc/9812002",
    archivePrefix = "arXiv",
    doi = "10.1103/PhysRevLett.81.4293",
    journal = "Phys. Rev. Lett.",
    volume = "81",
    pages = "4293",
    year = "1998"
}

@article{Parikh:1999mf,
    author = "Parikh, Maulik K. and Wilczek, Frank",
    title = "{Hawking radiation as tunneling}",
    eprint = "hep-th/9907001",
    archivePrefix = "arXiv",
    reportNumber = "PUPT-1775, SPIN-1998-12, IASSNS-HEP-98-22",
    doi = "10.1103/PhysRevLett.85.5042",
    journal = "Phys. Rev. Lett.",
    volume = "85",
    pages = "5042--5045",
    year = "2000"
}

@article{Angheben:2005rm,
    author = "Angheben, Marco and Nadalini, Mario and Vanzo, Luciano and Zerbini, Sergio",
    title = "{Hawking radiation as tunneling for extremal and rotating black holes}",
    eprint = "hep-th/0503081",
    archivePrefix = "arXiv",
    doi = "10.1088/1126-6708/2005/05/014",
    journal = "JHEP",
    volume = "05",
    pages = "014",
    year = "2005"
}

@article{Witten:2024upt,
    author = "Witten, Edward",
    title = "{Introduction to black hole thermodynamics}",
    eprint = "2412.16795",
    archivePrefix = "arXiv",
    primaryClass = "hep-th",
    doi = "10.1140/epjp/s13360-025-06288-y",
    journal = "Eur. Phys. J. Plus",
    volume = "140",
    number = "5",
    pages = "430",
    year = "2025"
}

@article{Tiomno:1972dq,
    author = "Tiomno, J.",
    title = "{Maxwell equations in a spherically symmetric black-hole background and radiation by a radially moving charge}",
    reportNumber = "PRINT-72-2483",
    doi = "10.1007/BF02832806",
    journal = "Lett. Nuovo Cim.",
    volume = "5S2",
    pages = "851--855",
    year = "1972"
}

@article{Sanchez:1976fcl,
    author = "Sanchez, Norma G.",
    title = "{Scattering of scalar waves from a Schwarzschild black hole}",
    doi = "10.1063/1.522949",
    journal = "J. Math. Phys.",
    volume = "17",
    number = "5",
    pages = "688",
    year = "1976"
}

@article{Regge:1957td,
    author = "Regge, Tullio and Wheeler, John A.",
    title = "{Stability of a Schwarzschild singularity}",
    doi = "10.1103/PhysRev.108.1063",
    journal = "Phys. Rev.",
    volume = "108",
    pages = "1063--1069",
    year = "1957"
}

@book{Carroll:2004st,
    author = "Carroll, Sean M.",
    title = "{Spacetime and Geometry}: {An Introduction to General Relativity}",
    doi = "10.1017/9781108770385",
    isbn = "978-0-8053-8732-2, 978-1-108-48839-6, 978-1-108-77555-7",
    publisher = "Cambridge University Press",
    month = "7",
    year = "2019"
}

@book{Friedlander:2010eqa,
    author = "Friedlander, F. G.",
    title = "{The Wave Equation on a Curved Space-Time}",
    isbn = "978-0-521-13636-5",
    publisher = "Cambridge University Press",
    month = "3",
    year = "2010"
}

@article{Poisson:2011nh,
    author = "Poisson, Eric and Pound, Adam and Vega, Ian",
    title = "{The Motion of point particles in curved spacetime}",
    eprint = "1102.0529",
    archivePrefix = "arXiv",
    primaryClass = "gr-qc",
    doi = "10.12942/lrr-2011-7",
    journal = "Living Rev. Rel.",
    volume = "14",
    pages = "7",
    year = "2011"
}

@book{FaddeevYakubovskii2009,
  title     = {Lectures on Quantum Mechanics for Mathematics Students},
  author    = {L. D. Faddeev and O. A. Yakubovski{\v\i}},
  editor    = {Leon A. Takhtajan},
  series    = {Student Mathematical Library},
  volume    = {47},
  publisher = {American Mathematical Society},
  address   = {Providence, RI},
  year      = {2009},
  isbn      = {978-0-8218-4699-5},
  note      = {With an appendix by Leon Takhtajan},
}

@book{Landau:1991wop,
    author = "Landau, Lev Davidovich and Lifshits, E. M.",
    title = "{Quantum Mechanics}: {Non-Relativistic Theory}",
    doi = "10.1016/C2013-0-02793-4",
    isbn = "978-0-7506-3539-4",
    publisher = "Butterworth-Heinemann",
    address = "Oxford",
    series = "Course of Theoretical Physics",
    volume = "v.3",
    year = "1991"
}

@book{Newton:1982qc,
  title     = {Scattering Theory of Waves and Particles},
  author    = {Newton, Roger G.},
  edition   = {2nd},
  series    = {Texts and Monographs in Physics},
  year      = {1982},
  publisher = {Springer-Verlag},
  address   = {New York},
  isbn      = {978-0-387-10950-3}
}

@article{Newton1962,
  author  = {Roger G. Newton},
  title   = {Nonrelativistic S-Matrix Poles for Complex Angular Momenta},
  journal = {Journal of Mathematical Physics},
  volume  = {3},
  number  = {5},
  pages   = {867--879},
  year    = {1962},
  month   = sep,
  doi     = {10.1063/1.1724307}
}

@book{tong2025quantum,
  author    = {Tong, David},
  title     = {Quantum Mechanics},
  series    = {Lectures on Theoretical Physics},
  volume    = {3},
  publisher = {Cambridge University Press},
  year      = {2025},
  address   = {Cambridge},
  isbn      = {978-1009594820}
}

@book{Rakityansky2022,
  author    = {Sergei A. Rakityansky},
  title     = {Jost Functions in Quantum Mechanics},
  subtitle  = {A Unified Approach to Scattering, Bound, and Resonant State Problems},
  publisher = {Springer},
  address   = {Cham},
  year      = {2022},
  isbn      = {978-3-031-07760-9},
  doi       = {10.1007/978-3-031-07761-6},
  url       = {https://doi.org/10.1007/978-3-031-07761-6}
}

@book{whittaker1996course,
  title     = {A Course of Modern Analysis},
  author    = {Whittaker, E. T. and Watson, G. N.},
  editor    = {Moll, Victor H.},
  edition   = {5th},
  year      = {2021},
  publisher = {Cambridge University Press},
  address   = {Cambridge},
  isbn      = {978-1-316-51893-9},
  doi       = {10.1017/9781009004091}
}

@book{abramowitz1965handbook,
  title={Handbook of Mathematical Functions: With Formulas, Graphs, and Mathematical Tables},
  author={Abramowitz, Milton and Stegun, Irene A.},
  year={1965},
  publisher={Dover Publications},
  address={New York},
  isbn={978-0486612720}
}

@book{beals2010special,
  author    = {Beals, Richard and Wong, Roderick},
  title     = {Special Functions: A Graduate Text},
  series    = {Cambridge Studies in Advanced Mathematics},
  volume    = {126},
  publisher = {Cambridge University Press},
  year      = {2010},
  address   = {Cambridge},
  isbn      = {978-0521194341}
}

@misc{NIST:DLMF,
         key = "{\relax DLMF}",
       title = "{\it NIST Digital Library of Mathematical Functions}",
howpublished = "\url{https://dlmf.nist.gov/}, Release 1.2.6 of 2026-03-15",
         url = "https://dlmf.nist.gov/",
          year      = {2026},
        note = "F.~W.~J. Olver, A.~B. {Olde Daalhuis}, D.~W. Lozier, B.~I. Schneider,
                R.~F. Boisvert, C.~W. Clark, B.~R. Miller, B.~V. Saunders,
                H.~S. Cohl, and M.~A. McClain, eds."}

@book{squires1963complex,
  author    = {Euan J. Squires},
  title     = {Complex Angular Momenta and Particle Physics: A Lecture Note and Reprint Volume},
  series    = {Frontiers in Physics},
  volume    = {15},
  publisher = {W. A. Benjamin},
  address   = {New York},
  year      = {1963},
  pages      = {161}
}

@article{Gribov1962Regge,
  author    = {V. N. Gribov and I. Ya. Pomeranchuk},
  title     = {Regge Poles and Landau Singularities},
  journal   = {Physical Review Letters},
  volume    = {9},
  number    = {6},
  pages      = {238--240},
  year       = {1962},
  month      = sep,
  doi        = {10.1103/PhysRevLett.9.238},
  publisher  = {American Physical Society}
}

@article{Bekenstein:1972tm,
    author = "Bekenstein, J. D.",
    title = "{Black holes and the second law}",
    doi = "10.1007/BF02757029",
    journal = "Lett. Nuovo Cim.",
    volume = "4",
    pages = "737--740",
    year = "1972"
}

@article{Hartle:1976tp,
    author = "Hartle, J. B. and Hawking, S. W.",
    title = "{Path Integral Derivation of Black Hole Radiance}",
    doi = "10.1103/PhysRevD.13.2188",
    journal = "Phys. Rev. D",
    volume = "13",
    pages = "2188--2203",
    year = "1976"
}

@article{Bethe:1949yr,
    author = "Bethe, H. A.",
    title = "{Theory of the Effective Range in Nuclear Scattering}",
    doi = "10.1103/PhysRev.76.38",
    journal = "Phys. Rev.",
    volume = "76",
    pages = "38--50",
    year = "1949"
}

@article{BethePeierls1935,
  author  = {H. A. Bethe and R. Peierls},
  title   = {The Scattering of Neutrons by Protons},
  journal = {Proc. R. Soc. A},
  volume  = {149},
  number  = {866},
  pages   = {176--183},
  year    = {1935},
  doi     = {10.1098/rspa.1935.0055}
}

@book{Percacci:2017fkn,
    author = "Percacci, Robert",
    title = "{An Introduction to Covariant Quantum Gravity and Asymptotic Safety}",
    doi = "10.1142/10369",
    isbn = "978-981-320-717-2, 978-981-320-719-6",
    publisher = "World Scientific",
    series = "100 Years of General Relativity",
    volume = "3",
    year = "2017"
}

@article{tHooft:1974toh,
    author = "'t Hooft, Gerard and Veltman, M. J. G.",
    title = "{One-loop divergencies in the theory of gravitation}",
    doi = "10.1142/9789814539395_0001",
    journal = "Ann. Inst. H. Poincare Phys. Theor. A",
    volume = "20",
    number = "1",
    pages = "69--94",
    year = "1974"
}

@article{Goroff:1985th,
    author = "Goroff, Marc H. and Sagnotti, Augusto",
    title = "{The Ultraviolet Behavior of Einstein Gravity}",
    reportNumber = "CALT-68-1289, LBL-19995, UCB-PTH-85-34",
    doi = "10.1016/0550-3213(86)90193-8",
    journal = "Nucl. Phys. B",
    volume = "266",
    pages = "709--736",
    year = "1986"
}

@article{Deser:1974xq,
    author = "Deser, Stanley and Tsao, Hung-Sheng and van Nieuwenhuizen, P.",
    title = "{One Loop Divergences of the Einstein Yang-Mills System}",
    reportNumber = "Print-74-1164 (BRANDEIS)",
    doi = "10.1103/PhysRevD.10.3337",
    journal = "Phys. Rev. D",
    volume = "10",
    pages = "3337",
    year = "1974"
}

@article{Deser:1974cz,
    author = "Deser, Stanley and van Nieuwenhuizen, P.",
    title = "{One Loop Divergences of Quantized Einstein-Maxwell Fields}",
    reportNumber = "Print-74-0576 (BRANDEIS)",
    doi = "10.1103/PhysRevD.10.401",
    journal = "Phys. Rev. D",
    volume = "10",
    pages = "401",
    year = "1974"
}

@article{Deser:1974cy,
    author = "Deser, Stanley and van Nieuwenhuizen, P.",
    title = "{Nonrenormalizability of the Quantized Dirac-Einstein System}",
    reportNumber = "Print-74-0575 (BRANDEIS)",
    doi = "10.1103/PhysRevD.10.411",
    journal = "Phys. Rev. D",
    volume = "10",
    pages = "411",
    year = "1974"
}

@book{Schwartz:2014sze,
    author = "Schwartz, Matthew D.",
    title = "{Quantum Field Theory and the Standard Model}",
    doi = "10.1017/9781139540940",
    isbn = "978-1-107-03473-0, 978-1-107-03473-0",
    publisher = "Cambridge University Press",
    month = "3",
    year = "2014"
}

@book{Futterman:1988ni,
    author = "Futterman, J. A. H. and Handler, F. A. and Matzner, R. A.",
    title = "{Scattering From Black Holes}",
    doi = "10.1017/CBO9780511735615",
    isbn = "978-1-139-24539-5, 978-0-521-11210-9",
    publisher = "Cambridge University Press",
    series = "Cambridge Monographs on Mathematical Physics",
    month = "5",
    year = "2012"
}

@article{Mukhanov:1986me,
    author = "Mukhanov, Viatcheslav F.",
    title = "{ARE BLACK HOLES QUANTIZED?}",
    journal = "JETP Lett.",
    volume = "44",
    pages = "63--66",
    year = "1986"
}

@inproceedings{Bekenstein:1997bt,
    author = "Bekenstein, Jacob D.",
    title = "{Quantum black holes as atoms}",
    booktitle = "{8th Marcel Grossmann Meeting on Recent Developments in Theoretical and Experimental General Relativity, Gravitation and Relativistic Field Theories (MG 8)}",
    eprint = "gr-qc/9710076",
    archivePrefix = "arXiv",
    pages = "92--111",
    month = "6",
    year = "1997"
}

@article{Giddings:2015uzr,
    author = "Giddings, Steven B.",
    title = "{Hawking radiation, the Stefan{\textendash}Boltzmann law, and unitarization}",
    eprint = "1511.08221",
    archivePrefix = "arXiv",
    primaryClass = "hep-th",
    doi = "10.1016/j.physletb.2015.12.076",
    journal = "Phys. Lett. B",
    volume = "754",
    pages = "39--42",
    year = "2016"
}

@article{Dey:2017yez,
    author = "Dey, Ramit and Liberati, Stefano and Pranzetti, Daniele",
    title = "{The black hole quantum atmosphere}",
    eprint = "1701.06161",
    archivePrefix = "arXiv",
    primaryClass = "gr-qc",
    doi = "10.1016/j.physletb.2017.09.076",
    journal = "Phys. Lett. B",
    volume = "774",
    pages = "308--316",
    year = "2017"
}

@article{Hod:2016hdd,
    author = "Hod, Shahar",
    title = "{Hawking radiation and the Stefan{\textendash}Boltzmann law: The effective radius of the black-hole quantum atmosphere}",
    eprint = "1607.02510",
    archivePrefix = "arXiv",
    primaryClass = "gr-qc",
    doi = "10.1016/j.physletb.2016.03.071",
    journal = "Phys. Lett. B",
    volume = "757",
    pages = "121--124",
    year = "2016"
}

@article{Dey:2019ugf,
    author = "Dey, Ramit and Liberati, Stefano and Mirzaiyan, Zahra and Pranzetti, Daniele",
    title = "{Black hole quantum atmosphere for freely falling observers}",
    eprint = "1906.02958",
    archivePrefix = "arXiv",
    primaryClass = "gr-qc",
    doi = "10.1016/j.physletb.2019.134828",
    journal = "Phys. Lett. B",
    volume = "797",
    pages = "134828",
    year = "2019"
}

@article{Zhang:2026ixo,
    author = "Zhang, Shuai and Wang, Dong",
    title = "{Quantum resource theories in quantum atmosphere}",
    doi = "10.1016/j.physletb.2026.140354",
    journal = "Phys. Lett. B",
    volume = "875",
    pages = "140354",
    year = "2026"
}

@article{Liu:2026exs,
    author = "Liu, Xiaofang and Wen, Cuihong and Wang, Jieci",
    title = "{Quantum coherence of continuous variables in the black hole quantum atmosphere}",
    eprint = "2601.06741",
    archivePrefix = "arXiv",
    primaryClass = "gr-qc",
    doi = "10.1016/j.physletb.2026.140185",
    journal = "Phys. Lett. B",
    volume = "873",
    pages = "140185",
    year = "2026"
}

@article{Ong:2020hti,
    author = "Ong, Yen Chin and Good, Michael R. R.",
    title = {{Quantum atmosphere of Reissner-Nordstr{\"o}m black holes}},
    eprint = "2003.10429",
    archivePrefix = "arXiv",
    primaryClass = "gr-qc",
    doi = "10.1103/PhysRevResearch.2.033322",
    journal = "Phys. Rev. Res.",
    volume = "2",
    number = "3",
    pages = "033322",
    year = "2020"
}

@article{Kaczmarek:2024QuantumAtmosphere,
    author = "Kaczmarek, Adam Z. and Szczkesniak, Dominik",
    title = {{Signatures of the black hole quantum atmosphere in nonlocal correlations}},
    doi = "10.1016/j.physletb.2023.138364",
    journal = "Phys. Lett. B",
    volume = "848",
    pages = "138364",
    year = "2024"
}

@article{Gingrich:2023qae,
    author = "Gingrich, Douglas M.",
    title = {{Quantum atmosphere effective radii for different spin fields from quantum gravity inspired black holes}},
    doi = "10.1007/s10714-023-03131-6",
    journal = "Gen. Rel. Grav.",
    volume = "55",
    pages = "80",
    year = "2023"
}

@article{Berti:2005ys,
    author = "Berti, Emanuele and Cardoso, Vitor and Will, Clifford M.",
    title = "{On gravitational-wave spectroscopy of massive black holes with the space interferometer LISA}",
    eprint = "gr-qc/0512160",
    archivePrefix = "arXiv",
    doi = "10.1103/PhysRevD.73.064030",
    journal = "Phys. Rev. D",
    volume = "73",
    pages = "064030",
    year = "2006"
}

@article{biggs,
    author        = "Biggs, Anna and Maldacena, Juan",
    title         = "{Comparing the decoherence effects due to black holes versus ordinary matter}",
    eprint        = "2405.02227",
    journal = "arXiv",
    primaryClass  = "hep-th",
    month         = "5",
    year          = "2024"
}

@book{Wald:1995yp,
    author = "Wald, Robert M.",
    title = "{Quantum Field Theory in Curved Space-Time and Black Hole Thermodynamics}",
    isbn = "978-0-226-87027-4",
    publisher = "University of Chicago Press",
    address = "Chicago, IL",
    series = "Chicago Lectures in Physics",
    year = "1995"
}

@book{Birrell:1982ix,
    author = "Birrell, N. D. and Davies, P. C. W.",
    title = "{Quantum Fields in Curved Space}",
    doi = "10.1017/CBO9780511622632",
    isbn = "978-0-511-62263-2, 978-0-521-27858-4",
    publisher = "Cambridge University Press",
    address = "Cambridge, UK",
    series = "Cambridge Monographs on Mathematical Physics",
    year = "1982"
}

\end{document}